\documentclass[twocolumn,amsmath,amssymb,prl]{revtex4-1}

\usepackage{graphicx}% Include figure files
\usepackage{dcolumn}% Align table columns on decimal point
\usepackage{bm}% bold math

\begin{document}

\title{Three-state structural heterogeneity in a model two dimensional fluid}% Force line breaks with \\

\author{Tamoghna Das}
\email{dtamoghna@me.com}
\affiliation{
Center for Nanoscale Science and Technology, National Institute of Standards and Technology, Gaithersburg, MD 20899, USA
}
\affiliation{
Maryland Nanocenter, University of Maryland, College Park, MD 20742, USA.
}
\author{Jack F. Douglas}
\email{jack.douglas@nist.gov}
\affiliation{
Materials Science and Engineering Division, National Institute of Standards and Technology, Gaithersburg, Maryland 20899, USA.
}


\begin{abstract}
Three structural populations with distinct average mobility are identified within an equilibrium two-dimensional Lennard-Jones fluid simulated via molecular dynamics at a constant temperature and varying density. Quantifying the structure of the immediate neighborhood of particles by a tessellation allows us to identify three distinct structural states by the shapes of the tessellation cells. Irrespective of dynamic particle exchange among these populations, each is observed to maintain their own thermodynamic and average dynamic properties across the liquid-solid transition. We expect these findings to be valuable for better understanding the structural basis of dynamical heterogeneity in complex fluids defined in terms of local mobility fluctuations.

\end{abstract}

\maketitle

%\section{Introduction}
%{\it Dynamical heterogeneity} (DH)\cite{DHrev1,DHrev2} refers to localized spatial fluctuations in fluid properties that occur intermittently in a system. This term refers, in recent works,\cite{dw1,dw2} to the local {\em mobility} fluctuations.

It is now widely appreciated that localized fluctuations in mobility are characteristic features of strongly interacting dense fluids. \cite{dw1,dw2} Commonly referred as {\em dynamical heterogeneity} (DH), \cite{DHrev1,DHrev2} this phenomena is observed in virtually all condensed matter systems, \cite{DHbook} ranging from microscopic to macroscopic scales, such as colloidal dispersions, molecular and polymeric liquids, metallic alloys, plastics (thermal systems) and granular materials-like sands, powders, foams and pastes (athermal systems). While dynamically heterogeneity is usually described in terms of the local {\em mobility} fluctuations of the fluid molecules or particles, \cite{dw3} it is natural to expect some sort of complementary structural fluctuations that somehow facilitate such fluctuations, but the precise nature of these fluctuations is still  unresolved. \cite{DHrev3} The present work seeks to better quantify structural heterogeneity in a model strongly interacting dense fluids in order to better understand the origin of mobility fluctuations in this type of fluid.
%Serious attempts are underway to describe the DH solely in terms of the local {\em mobility} fluctuations of the constituents.
%Supports coming from recent observations of mobility fluctuations in mammalian cells, bacterial colony, collective animal movement etc. have further broadened the scope of this view to understand systems even beyond traditional materials.
%While it is natural to expect some sort of complementary structural fluctuations facilitating the fluctuations in mobility, precise nature of these fluctuations is still  unresolved. The present work seeks to better quantify the structural heterogeneity in a model strongly interacting dense fluids in order to better understand the origin of mobility fluctuations in this type of fluids.
%It is often imagined that the structural disorder allows for only short-range correlations in particle packing but the appearance of slow relaxation in the disordered materials requires 
%is thus attributable to the emergence of finite-range ordering over finite time a.k.a DH. However, a generic structural characterization of DH remains a major challenge to date.\cite{DHrev3}

The key question is how correlations in spatial structures influence the local dynamics of condensed materials.\cite{crr1,crr2,crr3,crr4,crr5} Over the years, various local fluid properties have been proposed for this purpose: free volume defined by local cavity in fluids, \cite{fv} local Debye-Waller factor as a measure of local stiffness, \cite{dw4} local elastic moduli, locally favored packing structures, \cite{lfs1,lfs2} local thermal energy, \cite{lte} to name a few proposed forms of structural heterogeneity. Some proposed structural measures are derived from the knowledge of particle position alone, while others require additional information about interatomic interaction and/or vibrational properties. While each of these local structural properties has had some success in understanding the dynamics of fluids, they often are system specific and lack consistency when compared with other structural indicators. Sometimes, the physical foundation of such indicators remains unspecified, as in recent application of machine learning approaches to glass-forming liquids. \cite{ml1,ml2}

The notion of DH was initially introduceed \cite{vsco} to help understand the dynamical response of glass-forming liquids. In such materials, the existence of particles belonging to clusters of finite size and lifetime implies the existence of spatio-temporal correlations of some sort within the system. Certain long-lived fluctuations are then required to disintegrate such structures to allow their flow through structural relaxation. This suggests the coexistence of at least two dynamic populations of {\em fast} and {\em slow} particles \cite{DHrev4} in this class of dynamically heterogeneous fluids. Similar broad distributions of particle mobility also arise in many other condensed materials at equilibrium, especially, near a liquid-solid phase transition. Given the stark difference in the long-time dynamics of liquids and crystalline media, a {\em critical} system undergoing such transition can easily be conceived as an admixture of dynamic populations with different mobility which shows average slow dynamics and increasing correlation length due to structural ordering, characteristic of DH. The obvious structural change across the liquid to solid transition also makes such structures suitable for understanding the structure-dynamics relations in a controlled setting other than the glass-forming liquids. 

To this end, we employ a {\em radial angular distance} (RAD) \cite{rad} based space tessellation method to characterize the structure of a two-dimensional (2D) model single component equilibrium fluid approaching a liquid-solid phase transition by increasing density at a fixed temperature. This newly introduced technique is argued to better capture the local coordination of particles within a inhomogeneous polydispers system than the Voronoi tessellation. A 2D case is considered for the sake of simplicity and easy visualization. The local structures, quantified in terms of the shape of first coordination shells, then reveal at least {\em three} dynamic population contributing to DH which are structurally, thermodynamically and dynamically distinct from each other. 
%Possible applications of present method and its extension to higher dimensions is also discussed.
%To this end, this work presents a geometric characterization of structures of a model single component equilibrium fluid approaching a liquid-solid phase transition by increasing density at a fixed temperature. A two-dimensional case is considered for the sake of simplicity and easy visualization. A solid angle based tessellation technique is used to identify the coordination shell of each particles in simulated configurations. The local structure quantified in terms of the shape of such shells then reveals {\em three} dynamic population contributing to DH which are structurally different from each other. Possible applications of present method and its extension to higher dimensions is also discussed.

%The local structure of this single component equilibrium system is quantified in terms of the shape of the nearest neighborhood of constituent particles identified by a solid angle based method. Most importantly, this characterization naturally identifies three dynamic population contributing to DH which are structurally different from each other. Possible applications of present method and its extension to higher dimensions is also discussed.

%\section{Materials and Methods}
{\em Model System and Simulation Details --}
We use large-scale molecular dynamics simulation \cite{mdbook} to prepare a two-dimensional equilibrium system of about $60,000$ particles in a canonical constant number-area-temperature (NAT) ensemble. Particles of size $\sigma$ and unit mass interact via a Lennard-Jones potential $V(r) = 4\epsilon [(\sigma/r)^{12}-(\sigma/r)^6]$ where $\epsilon$ is unit energy and $r$ is the distance between a pair of particles. Temperature $k_BT=0.7$ (in $\epsilon$ unit) is fixed by a {\em DPD} \cite{dpd} thermostat as implemented within {\em LAMMPS} \cite{lammps} where $k_B$ is the Boltzmann constant set to unity. Measuring time in units of $\tau=\sqrt{m\sigma^2/\epsilon}$, numerical integration of the equations of motion is done in steps of $\delta t=10^{-3}$ with a velocity Verlet algorithm. Systems with $30$ different $\rho$, densities are considered ranging from $\rho=0.7$ to $1.0$ within a square box of suitable size with periodic boundary conditions in all directions. After equilibrating each system over $10^3 \tau$, particle trajectories of length $100 \tau$ are stored for analysis. Equilibration is confirmed by the total energy fluctuation which is of the order of $10^{-3}$ over the observation period. Further details of the simulation and thermodynamic characterization of the system is presented elsewhere in detail.\cite{p2} For the choice of densities, the model system shows \cite{2dLJ} liquid, crystalline and liquid-solid coexistence behavior at $T=0.7$. Below, we present and discuss the main findings of our study.

\begin{figure}[h!]
\begin{center}
\includegraphics[width=0.95\linewidth]{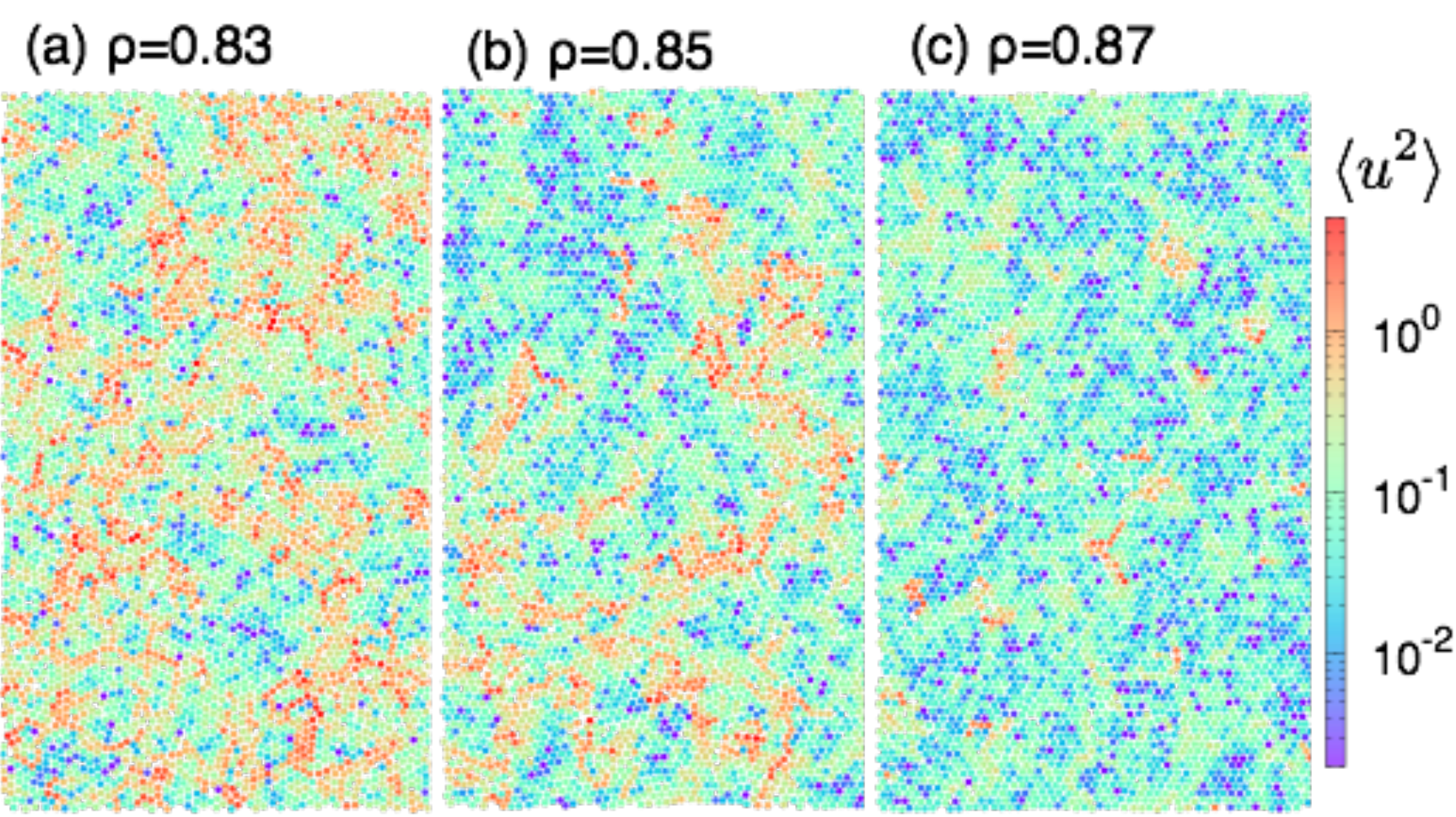}
\end{center}
\caption{
(color online) Spatial map of mobility $\langle u^2\rangle$ (defined in text) is presented for three different densities (a) $\rho=0.83$, (b) $\rho=0.85$ and (c) $\rho=0.87$. Only a part of the simulation box is shown in each cases. Note that $\langle u^2\rangle$ for {\em red} (highly mobile) and {\em blue} (nearly immobile) particles differ by nearly $4$-orders of magnitude. The {\em string-like} structure of particles having a relatively high mobility is a conspicuous feature of glass-forming liquids and many strongly strongly interacting particle systems.
%Signature {\em string-like} motion of particles with similar mobility provides evidence of heterogeneous dynamics in all the densities.
}
\label{mobility}
\end{figure}

%\section{Results}
%\subsection{DH in equilibrium}
{\em Equilibrium Dynamic Heterogeneity --}
Following the simulated trajectories of particles, it is easy to quantify their relative {\em mobility} in terms of their displacement over certain time period. We define individual particle mobility as $u_i^2=|{\bf r}_i(\tau)-{\bf r}_i(0)|^2$ where ${\bf r}_i(t)$ is the position vector of $i$-th particle at time $t$. Although the choice of time period $\tau$ is somewhat arbitrary, it is long enough to observe overall diffusive behavior in our model system. Color-coding the instantaneous particle configuration for $t>\tau$ immediately shows (Fig.\ref{mobility}) the presence of fast ({\em red}) and slow ({\em blue}) particles with mobilities differing by the orders of magnitude for all densities considered. We see that the number of highly mobile particles decreases as particle crowding occurs with the increase in density and more particles become immobile. Importantly, the spatial organization of particles with similar mobility become apparent from such visualization which reveal the characteristic {\em string-like} arrangement of the most mobile particles,\cite{string1,string2} a hallmark of DH. Now that we have confirmed the existance of DH in our equilibrium model system, we focus on quantifying the structural heterogeneity of the same fluid.

%\subsection{Geometry of neighborhood}
{\em Geometrical Neighborhood as a Measure of Structural Dynamical Heterogeneity --}
Identification of the immediate neighborhood of individual particles is a natural first step to gather structural information about a disordered system. As such systems enjoy the full translational and rotational symmetry, obtaining complete knowledge of both radial and angular arrangement of particles around a central one is a difficult task and no unique method to do so has been prescribed so far. Two widely used methods in this regard are the radial distribution function (RDF) based method and Voronoi tessellation (VT) technique. \cite{vt0,vt1} RDF based methods suffer from the arbitrariness of predefined cutoff to identify the closest neighbors. Also, this method which involves ensemble averaging over angles and configurations, is only capable of providing average properties relating to fluctuations in the particle arrangement so that this method does not provide insight into the dynamics of a particular particle arrangement. The VT provides more detailed information of particles' neighborhood by defining space-filling non-overlapping polygons around each of the particles. The area of such polygons in two dimensions and volumes in three dimensions are known to follow a well-defined distributions, \cite{vt2} universal to a wide variety of systems such as glass-forming liquids,\cite{vt3} granular materials, \cite{vt4,vt5} etc. However, it is now well appreciated that free volume defined by this local measure of local packing is inadequate for predicting local dynamics. \cite{fvdh1, fvdh2} It is possible that other measures of local structures based on the Voronoi construction, such as, number of faces or edges of the Voronoi cells, average shape of the cells, etc. might be more informative about local dynamics. There is the additional problem that the Voronoi based methods are also computationally expensive and often require further modifications for highly inhomogeneous materials such as aggregate forming systems, polymer liquids etc . The matter of computational efficiency motivated us to consider another tessellation method, radial angular displacement (RAD) \cite{rad} to quantify the geometrical neighborhood of finite-size particles in any arrangement. This tessellation is particularly good at quantifying changes in local coordination number which we intutively expect to correlate to changes in local fluid dynamics since this physical property should alter the local cohesive potential energy and local barrier for molecular transport.
%We, thus, adapt another geometric method which employs RAD \cite{rad} to identify nearest neighbors of a finite-size particle in any arrangement. 

The RAD method identifies a particle $j$ as nearest neighbor to particle $i$ if it is not blocked by any other particle $k$ and not further away from any other blocked particle. The blocking condition is ensured by the following geometric inequality: $\Omega_{ij}>\Omega_{ik}\theta_{jik}$ where $\theta_{jik}$ is the three-body angle made by $j$ and $k$ at $i$ and $\Omega_{ij}$ is the solid angle subtended by particle $j$ at the center of particle $i$. Making a small angle approximation for $\Omega_{ij}$, the criterion for equal size particles reads $r_{ik}^2>r_{ij}^2\theta_{jik}$, $r_{ij}$ is the distance between particles $i$ and $j$ and similarly, $r_{ik}$. Starting from a list of possible neighbors arranged in an order of ascending distance from the central particle, this parameter free method can quickly identify the first neighbor shells of all particles in a considerably large system as ours.

\begin{figure}[h!]
\begin{center}
\includegraphics[width=0.95\linewidth]{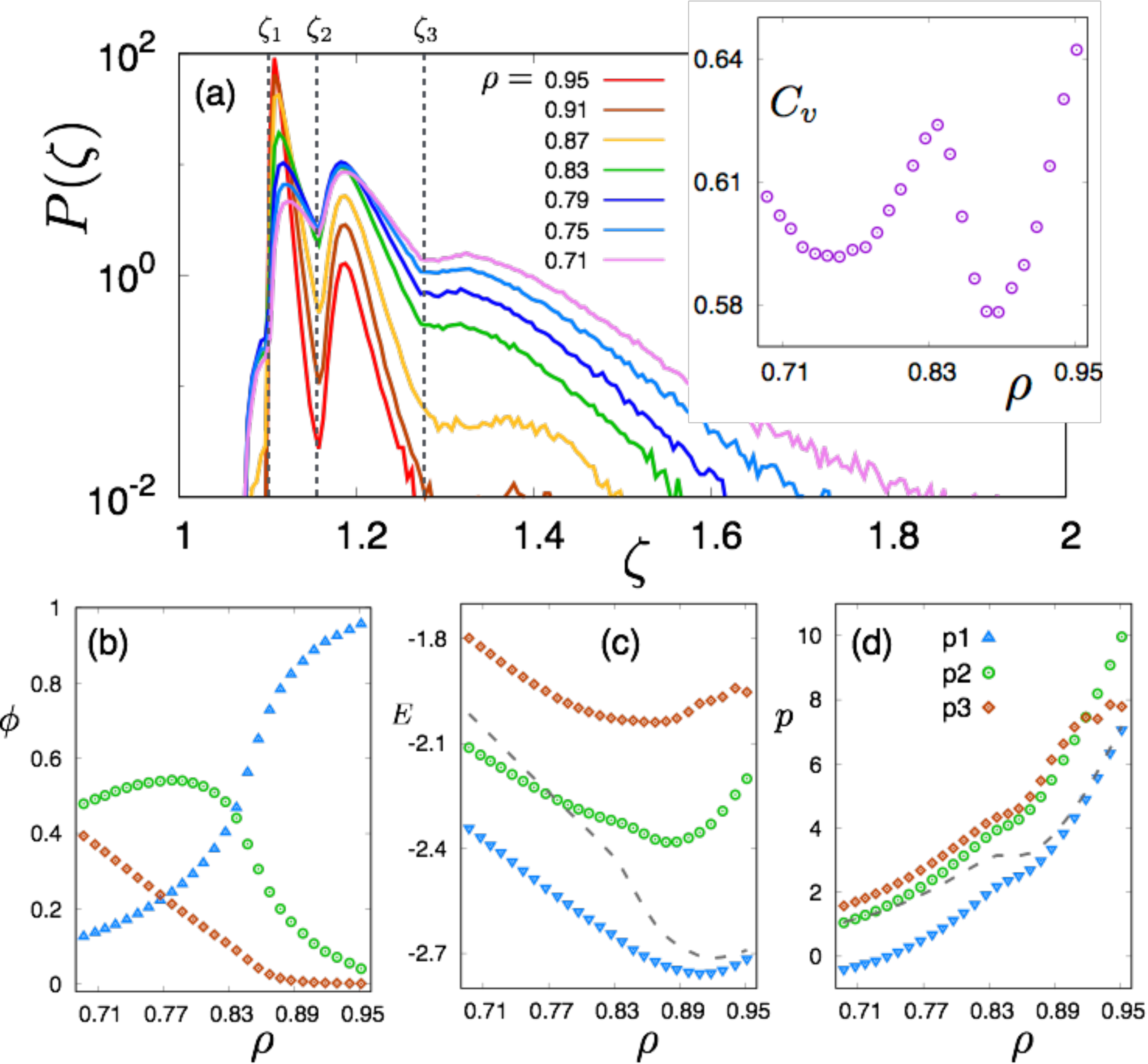}
\end{center}
\caption{
(color online) (a) The distribution of the shape factor $\zeta$ (defined in text) for the first nearest neighbor shell is plotted for different densities. {\em Inset} shows the phase transition points across these densities in terms of constant volume specific heat $C_v$. Systems with $\rho\le0.83$ behave as equilibrium liquids and crystalline behavior is observed for $\rho\ge0.89$ followed by intermediate densities with liquid-solid coexistence. $P(\zeta)$ shows two distinct minima at $\zeta_2$ and $\zeta_3$ for all densities. Based on this, the whole system can be imagined as an admixture of three distinct geometric populations:(i) $p_1$ $(\zeta<\zeta_2)$, (ii) $p_2$ $(\zeta_2\le\zeta\le\zeta_3)$ and (iii) $p_3$ $(\zeta>\zeta_3)$. (b) The change in the number density $\phi$ of each population is presented as a function of $\rho$. $p_1$ population shows a typical order parameter like behavior across liquid to solid transition, consistent with $C_v$. The other two populations behave in a complementary fashion. (c) and (d) show the distinct thermodynamic behavior of these three populations in terms of energy per particle, $E$ and virial pressure, $p$, respectively as a function of $\rho$. In each case, the {\em dashed} line shows the same respective quantities computed for the full systems.
}
\label{shape}
\end{figure}
The geometry of each tessellation cell is then quantified by its {\em shape factor}, $\zeta=\mathbf{C}^2/(4\pi\mathbf{S})$ \cite{sf0} where $\mathbf{C,S}$ are the circumference and area of the shell, respectively. We note that $\zeta$ is the inverse square root of a well-known shape descriptor, {\em circularity}. \cite{sf1} Variations of basically the same shape descriptor have already been used in previous morphometric studies \cite{sf2,sf3} under different names such as {\em reduced area}, \cite{sf4} {\em isoperimetric ratio}, \cite{sf5} etc. While $\zeta$ is designed to be unity for a circle, the analytic values for a regular $n$-gon can be obtained by $\zeta=(n/\pi)\tan(\pi/n)$. For example, perfect hexagon, pentagon and square has values $\zeta_1\approx1.103$, $\zeta_2\approx1.156$ and $\zeta_3\approx1.273$, respectively. The distribution $P(\zeta)$ shows a trimodal distribution with fixed minima at $\zeta_2$ and $\zeta_3$ (Fig.\ref{shape}(a)) for all $\rho$ studied. We note that this trimodal feature resulting from the RAD algorithm used is in contrast with the VT method. $P(\zeta)$ computed using Voronoi neighborhoods is reported to show a bimodal to unimodal transition across the phase transition in model hard-disk fluid \cite{sf0} granular media \cite{sfvor2} and colloidal systems. \cite{sfvor1} The existence of three structural populations, proposed in these works based on physical intuition, can now be clearly identified using our tessellation method. The significance of these peaks for thermodynamic and dynamical properties is considered below.

{\em Definitions of Three Structural Populations --}
%While perfect five- and four-fold structures are not allowed in two-dimensions, they seem to be present in the system with certain degree of distortion. 
The prominent peak close to $\zeta_1$ at high densities assures crystalline packing with expected hexagonal symmetry. With decreasing density, this peak shifts and broadens as distorted hexagons start to appear allowing more disorder in the system. Perfect five- and four-fold symmetries are not allowed in two dimensions which is consistent with the position of fixed minima in $P(\zeta)$. Particles with $\zeta_2\le\zeta\le\zeta_3$ and $\zeta>\zeta_3$ can then be statistically identified as distorted pentagons and squares with increasing number as $\rho$ decreases. This well-behaved trimodal distribution allows us to unambiguously identify three distinct geometric populations: (i) $p_1$ $(\zeta<\zeta_2)$, (ii) $p_2$ $(\zeta_2\le\zeta\le\zeta_3)$ and (iii) $p_3$ $(\zeta>\zeta_3)$. Similar structural characterization is also possible based on the area of local neighborhood as discussed elsewhere. \cite{p2} The shape factor criterion employed in this work is more rigorous as the differentiation of populations is based on well-defined analytical values rather than relying on the numerical fitting of the local area distribution adapted in the other method. Nonetheless, the structural populations identified by either of the criteria are essentially the same.
%Average static and dynamic properties of the three populations are presented next.

%\subsection{Thermodynamics of local geometry}
{\em Thermodynamics of Three Interconverting Structural States --}
The number density $\phi$ of each geometric population computed as a ratio of the number of particles within a population over the total number of particles in the system is shown in Fig.\ref{shape}(b). We view $\phi$ for $p_1$ population as being a kind of ``order parameter" that exhibits a sigmoidal variation from a very low value at low $\rho$ and to unity at high $\rho$. A complementary variation is observed for $p_2$ which has a finite value at low $\rho$ and decreases in a sigmoidal way to a very small positive value with increasing $\rho$. Interestingly, $\phi\approx0.5$ for both $p_1$ and $p_2$ near $\rho=0.83$ which marks the onset of liquid-solid coexistence phase, as signaled by a prominent inflection point in $C_v$, the specific heat at constant volume computed from the potential energy fluctuation as a function of $\rho$. ({\em Inset} Fig.\ref{shape}(a)) We see that $\phi$ for $p_3$ is negligible at high $\rho$, but we find a linear increase of $\phi$ with decreasing $\rho$ starting from $\rho=0.88$, where we observe another sharp inflection point in $C_v$. A significant relation between local geometry and thermodynamics of the system is indicated as we observe that all three populations show well-separated mean energy values E at all $\rho$. (Fig.\ref{shape}(c)) We mention that the number of particles $n$ involved in computation of $E=\sum_iV(r_i)/n$ is not restricted to just the nearest neighbors but includes all particles within $2.5\sigma$, a cut-off set by the potential. Similarly, the hydrostatic pressure $p$ computed from the trace of virial stress tensor of each population is different from each other for all $\rho$ (see Fig.\ref{shape}(d)). The pressure $p$ for different populations plotted together with the bulk pressure of the system (shown by dashed line in Fig.\ref{shape}(d)) indicates that the overall thermodynamics for high $\rho$ systems is dominated by $p_1$ population, particles with hexagon-like neighborhood while for low $\rho$ systems with mostly pentagonal neighborhood i.e. $p_2$ population, consistent with previous observations of $\phi$. Particles within the $p_3$ population, having square-like neighborhoods represent locally high pressure regions as their average pressure is always larger than the bulk. The well-defined differences in thermodynamic properties among these populations of particles naturally lead to a consideration of the time evolution of these classes of particles.
%Clear thermodynamic distinction among the populations set the necessary physical basis of our geometric analysis of simulated equilibrium configurations.

%\subsection{Time evolution of local geometry and DH}
{\em Time Evolution of Structural State Populations --}
\begin{figure}[h!]
\begin{center}
\includegraphics[width=0.95\linewidth]{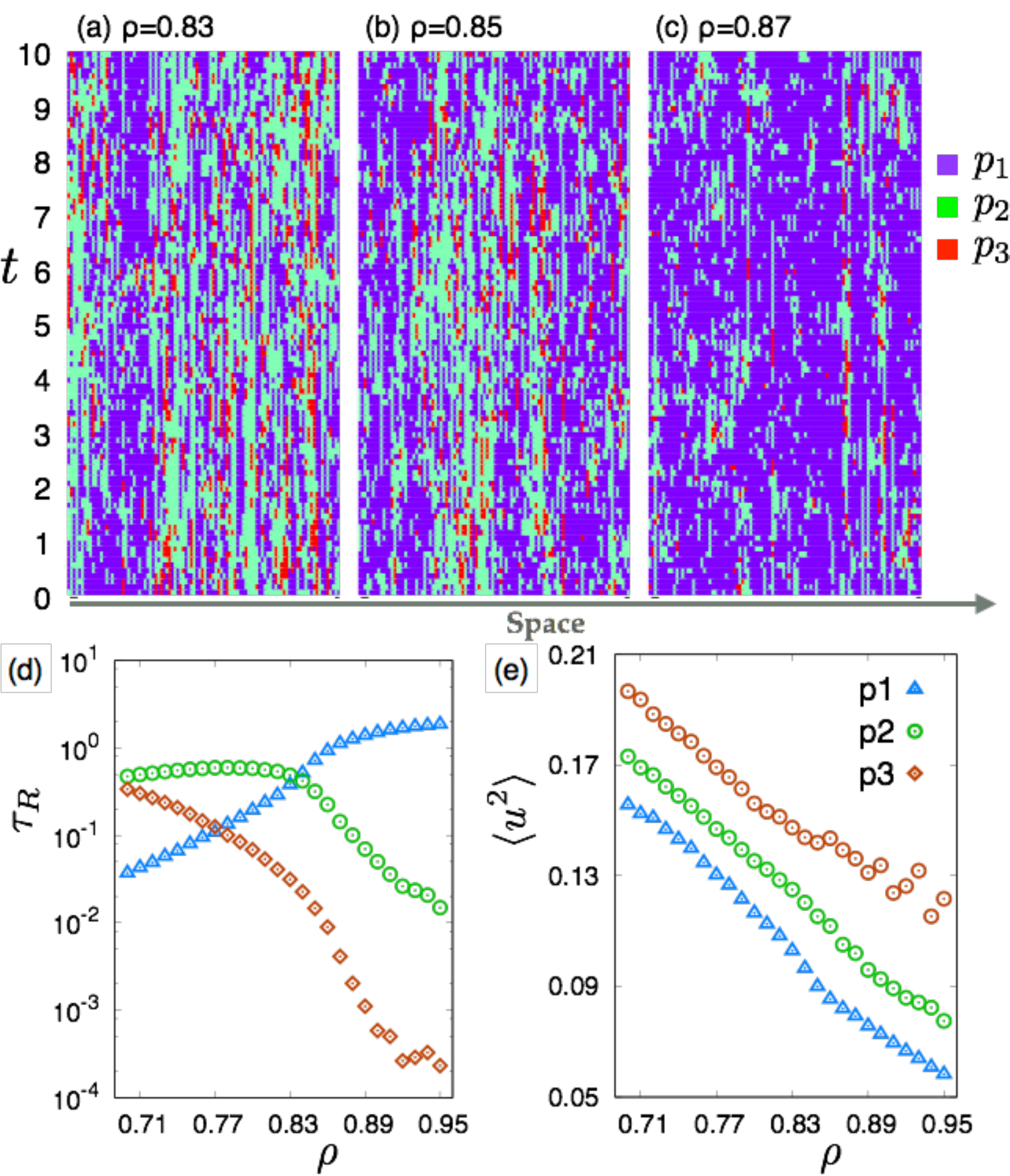}
\end{center}
\caption{
(color online) Visualization of the geometric fluctuations over time is presented for three different densities (a) $\rho=0.83$, (b) $\rho=0.85$ and (c) $\rho=0.87$. For each case, particles along a linear direction within the system is chosen arbitrarily. Each particle is initially colored according to their geometric identity, $p_1$ ({\em blue}), $p_2$ ({\em green}) and $p_3$ ({\em red}). As time progresses, particles retain their color if their identity remains the same and they acquire a new color when their neighborhood geometry is changed. This visualization gives us an idea about the lifetime of each type of populations.  (d) Average lifetime $\tau_R$ of the local geometry of an individual as a function of density is consistent with previous finding. Small $\tau_R$ means a mobile environment compared to larger $\tau_R$. (e) Average mobility $\langle u^2\rangle$ for individual populations is plotted as a function of density. $p_1$ is the least mobile, $p_3$ is most mobile and $p_2$ has an intermediate mobility for all $\rho$. This is a central result of the present study.
}
\label{hetero}
\end{figure}
We first attempt to visualize the local geometry as it evolves in time. First, we consider a set of particles chosen along a line cross-sectional cut of Fig.\ref{mobility} where the particles on this line are colored at each time step according to their population identity: {\em blue} ($p_1$), {\em green} ($p_2$) and {\em red} ($p_3$). A line segment of same color then represents the time period over which the shape of the neighborhood of an individual particle is maintained. These kymographs [REF.] are presented for three different $\rho$: Fig.\ref{hetero}(a) $0.83$, (b) $0.85$ and (c) $0.87$. At low density, the neighborhood changes its shape quite frequently. However, the population identity of adjacent particles seem to maintain some degree of correlation at least for $p_1$ and $p_2$ populations. The persistence of the {\em blue} lines increases with increasing $\rho$ pointing, that the $p_1$ particles maintains their neighborhood shape much longer than other populations. This is expected since the hexagonal shaped cells, characteristic of $p_1$, emerge in the system with increasing $\rho$. From this exercise, we can now quantify the average life time $\tau_R$ of the local geometry around a particle. We find that each population generally has different $\tau_R$ at a fixed density.(Fig.\ref{hetero}(d)) For high $\rho$ (crystalline phase), characteristic $\tau_R$ for each population is well separated by values of different order of magnitudes. $\tau_R$ as a function $\rho$ follows the same behavior of $\phi$ which is expected. We note that small $\tau_R$ is indicative of a fast changing local environment than a particle with larger $\tau_R$. This brings us back to the question of DH in the mobility sense. A calculation of the mobility $\langle u^2\rangle$ computed separately for each population, correspondingly shows distinctly different values at each $\rho$.(Fig.\ref{hetero}(e)) Evidently, $p_1$ and $p_3$ correspond to the immobile and highly mobile particle populations across all $\rho$, while $p_2$ maintains an intermediate mobility. Our geometric characterization of structure is then successful in identifying the key structural population across the phase transition.

%\section{Concluding remarks}
{\em Concluding Remarks --}
In summary, a scale- and parameter-free analysis for simulated particle configurations of a model two-dimensional equilibrium liquid is presented. We emphasize that both the identification of neighborhood and its shape analysis are crucial to recognize the key structural motifs with distinctly different thermodynamics and dynamical properties. Since both of the methods have already been applied to the three dimensional (3D) systems separately, their combination as in our case should easily be translated to 3D systems. In fact, as a purely geometric approach which uses only the positional information of the constituents of a system, this method should be readily applicable to practically any inhomogeneous arrangement of constituent particles, even those that have significant polydispersity in 3D. Being relatively computationally efficient, our method can be used for on-the-fly structural characterization of equilibrium systems and non-equilibrium fluids like granular media, colloidal suspensions, nanoparticle composites etc. We note that competition in three population systems have also been intensively studied in the context of population dynamics of living systems to explain the spatial organization and movement of interacting species.\cite{strngLV,strngEco} Although the structural populations revealed by our study are dynamical interchangeable and thus, different from those studied in population dynamics, constructing a model dynamical systems theory by taking inputs from our study can be instructive. We believe that our approach to defining structural heterogeneity in strongly interacting fluids provides a promising first step forward in development of a rigorous structure-dynamics relation that should be applicable to a broad class of condensed matter systems. However, further analysis of the model will be required to demonstrate the generality and practical utility of this relationship.%Taken together, the approach presented in this study can be an important step forward to a rigorous structure-dynamics relation of a broad class of condensed matter systems which is long awaiting.

{\em Acknowledgement} TD acknowledges support under the Cooperative Research Agreement between the University of Maryland and the National Institute of Standards and Technology Center for Nanoscale Science and Technology, Award 70NANB10H193, through the University of Maryland.

Official work of the US Government- Not subject to copyright in United States.


\begin{thebibliography}{}
%\bibitem{LiquidBook} J.-P. Hansen and I. R. McDonald, {\it Theory of Simple Liquids}, Academic Press, (2006).
%\bibitem{SolidBook} C. Kittel, {\it Introduction to Solid State Physics}, Wiley, (2004).
\bibitem{dw1} L. Larini, A. Ottochian, C. de Michele and D. Leporini,  {\it Nat. Phys.} {\bf 4}, 42 (2008).
\bibitem{dw2} G. M. Hocky, D. Coslovich, A. Ikeda and D. R. Reichman, {\it Phys. Rev. Lett.} {bf 113}, 157801 (2014).
\bibitem{DHrev1} M. D. Ediger, {\it Annu. Rev. Phys. Chem.} {\bf 51}, 99 (2000).
\bibitem{DHrev2} L. Berthier and G. Biroli, {\it Rev. Mod. Phys.} {\bf 83}, 587 (2011).
\bibitem{DHbook} L. Berthier, G. Biroli, J.-P. Bouchod, L. Cipellitti and W. van Saarloos, {\em Dynamical Heterogenities in Glasses, Colloids and Granular Media} (Oxford University Press, New York, 2011).
\bibitem{dw3} B. A. Pazmi\~{n}o Betancourt, P. A. Hanakata, F. W. Starr and J. F. Douglas, {\it Proc. Nat. Acad. Sc.} {\bf 112}, 1418654 (2015).
\bibitem{DHrev3} C. P. Royall and S. R. Williams, {\it Phys. Rep.} {\bf 560}, 1-75 (2015).
\bibitem{crr1} G. Adam and J. H. Gibbs, {\it J. Chem. Phys.} {\bf 43}, 139 (1965).
\bibitem{crr2} R. W. Hall and P.G. Wolynes, {\it J. Chem. Phys.} {\bf 86}, 2943 (1987).
\bibitem{crr3} P. G. Debenedetti and F. H. Stillinger, {\it Nature} {\bf 410} 259 (2001).
\bibitem{crr4} F. W. Starr, J. F. Douglas and S. Sastry, {\it J. Chem. Phys.} {\bf 138}, 12A541 (2013).
\bibitem{crr5} K. F. Freed, {\it J. Chem. Phys.} {\bf 141}, 141102 (2014).
\bibitem{fv} F. Spaepen, {\it Scr Mater} {\bf 54}, 363 (2006).
%\bibitem{cge1} G. S. Matharoo, M. S. G. Razul and P. H. Poole, {\it Phys. Rev. E} {\bf 74}, 050502 (2006).
%\bibitem{cge2} L. Berthier and R. L. Jack, {\it Phys. Rev. E} {\bf 76}, 041509 (2007).
\bibitem{dw4} A. Widmer-Cooper and P. Harowell, {\it Phys. Rev. Lett.} {\bf 96}, 185701 (2006).
\bibitem{lfs1} D. Coslovich and G. Pastore, {\it J. Chem. Phys.} {\bf 127}, 124504 (2007).
\bibitem{lfs2} C. P. Royall, S. R. Williams, T. Ohtsuka and H. Tanaka, {\it Nat. Mater.} {\bf 7}, 556 (2008).
%\bibitem{lem} M. Tsamados, A. Tanguy, C. Goldenberg and J. L. Barrat, {\it Phys. Rev. E} {\bf 80}, 026112 (2009).
\bibitem{lte} J. Zylberg, E. Lerner, Y. Bar-Sinai and E. Bouchbinder, {\it Proc. Nat. Acad. Sc.} {\bf 114}, 7289 (2017).
\bibitem{ml1} E. D. Cubuk, S. S. Schoenholz, J. M. Rieser, B. D. Malone, J. Rotler, D. J. Durian, E. Kaxiras, A. J. Liu, {\it Phys. Rev. Lett.} {\bf 114} 108001 (2015).
\bibitem{ml2} S. S. Schoenholz, E. D. Cubuk, D. M. Sussman, E. Kaxiras, A. J. Liu, {\it Nat. Phys} {\bf 12} 469 (2016).
\bibitem{vsco} J. H. Hildebrand, {\it Proc. Nat. Acad. Sc.} {\bf 72} 1970, (1975).
\bibitem{DHrev4} H. Sillescu, {\it J. Non-Cryst. Solids} {\bf 243}, 81 (1999).
\bibitem{rad} J. Higham and R. H. Henchman, {\it J. Chem. Phys.} {\bf 145} 084108 (2016).
%\bibitem{nodescrpzn} F. Spaepen, {\it Nature} {\bf 408} 781, (2000).
\bibitem{mdbook} D. Frenkel and B. Smit, {\em Understanding Molecular Simulation} (Academic Press, 1996).
\bibitem{dpd} R. D. Groot and P. B. Warren, {\it J. Chem. Phys.} {\bf 107}, 4423 (1997).
\bibitem{lammps} Freely available at {\texttt www.lammps.sandia.gov}
\bibitem{p2} T. Das and J. F. Douglas, ({\em to be published}).
\bibitem{2dLJ} J. A. Barker, D. Henderson and F. F. Abraham, {\it Physica} {\bf 106A}, 226 (1981).
\bibitem{string1} H. Zhang, P. Kalvapalle and J. F. Douglas, {\it Soft Matter} {\bf 6}, 5944 (2010).
\bibitem{string2} B. A. Pazmi\~{n}o Betancourt, J. F. Douglas and F. W. Starr, {\it Soft Matter} {\bf 9}, 241 (2012).
\bibitem{vt0} G. Voronoi and J. Reine, {\it Angew. Math.} {\bf 134}, 198 (1908).
\bibitem{vt1} Marina L. Gavrilova (Ed.), {\em Generalized Voronoi Diagram: A Geometry-Based Approach to Computational Intelligence} (Springer-Verlag Berlin Heidelberg, 2008).
\bibitem{vt2} V. S. Kumar and V. Kumaran, {\it J. Chem. Phys.} {\bf 123}, 114501 (2005).
\bibitem{vt3} F. W. Starr, S. Sastry, J. F. Douglas and S. C. Glotzer, {\it Phys. Rev. Lett.} {\bf 89}, 125501 (2002).
\bibitem{vt4} T. Aste, T. D. Matteo, M. Saadatfar, T. J. Senden, M. Schr\"{o}ter and H. L. Swinney, {\it Europhys. Lett.} {\bf 79}, 24003 (2007).
\bibitem{vt5} S. Slotterback, M. Toiya, L. Goff, J. F. Douglas and W. Losert, {\it Phys. Rev. Lett.} {\bf 101}, 258001 (2008).
\bibitem{fvdh1} F. W. Starr, S. Sastry, J. F. Douglas, and S. C. Glotzer, {\it Phys. Rev. Lett.} {\bf 89}, 125501(2002).
\bibitem{fvdh2} A. Widmer-Cooper and P. Harrowell, {\it J. Non-cryst. Solids} {\bf 352} 5098 (2006).
\bibitem{sf0} F. Mou\v{c}ka and I. Nezbeda, {\it Phys. Rev. Lett.} {\bf 94}, 040601 (2005).
\bibitem{sf1} A. M. Bouwman, J. C. Bosma, P. Vonk, J. A. Wesselingh, H. W. Frijlink, {\it Powder Technol.} {\bf 146}, 66 (2004).
\bibitem{sf2} J. Wasen, R. Warren, {\it Mater. Sci. Tech.} {\bf 5}, 222 (1989).
\bibitem{sf3} R. McAfee, I. Nettleship, {\it Acta Mater.} {\bf 51}, 4603 (2003).
\bibitem{sf4} A. Ho\v{c}evar, S. El Shawish and P. Ziherl, {\it Eur. Phys. J. E} {\bf 33}, 369 (2010).
\bibitem{sf5} V. Lucarini, {\it Symmetry} {\bf 1}, 21 (2009).
\bibitem{sfvor1} P. M. Reis, R. A. Ingale and M. D. Shattuck, {\it Phys. Rev. Lett.} {\bf 96}, 258001 (2006).
\bibitem{sfvor2} Z. Wang, A. M. Alsayed, A. G. Yodh and Y. Han, {\it J. Chem. Phys.} {\bf 132}, 154501 (2010).
\bibitem{strngLV} P. P. Avelino, D. Bazeia, J. Menezes and B. F. Oliveira, {\it Physics Letters A} {\bf 378}, 393 (2014).
\bibitem{strngEco} P. P. Avelino, D. Bazeia, L. Losano, J. Menezes and B. F. Oliveira, {\it Physics Letters A} {\bf 381}, 1014 (2017).

\end{thebibliography}
\end{document}